\documentstyle[12pt]{article}
\begin{document}
\title{WEYL'S LAW WITH ERROR ESTIMATE}
\author{Sultan Catto$^1$\\
{\it Physics Department}\\
{\it Baruch College and The Graduate School and University Center}\\
{\it 17 Lexington Avenue}\\
{\it New York, NY 10010}\\
{\it and}\\
{\it Physics Department, The Rockefeller University}\\
{\it 1230 York Avenue, New York, NY 10021-6399}\\
{\it cattos@rockvax.rockefeller.edu}\\ \\  \\  \\
Jonathan Huntley$^2$, Nam Jong Moh and David Tepper$^3$\\
{\it Math ematics Department, Baruch College}\\
{\it New York, NY 10010}\\
{\it huntley@gursey.baruch.cuny.edu}\\
{\it nj@gursey.baruch.cuny.edu}\\
{\it tepper@gursey.baruch.cuny.edu}}

\date{}
\maketitle

\newpage
\begin{abstract}
Let 
$$
X=Sl(3,Z)\backslash Sl(3,R)/SO(3,R). \eqno{}
$$
Let $N(\lambda )$ denote the dimension of the space of cusp forms with
Laplace eigenvalue less than $\lambda $. We prove that 
$$
N(\lambda )=C\lambda ^{\left( \frac 52\right) }+O(\lambda ^2)
\eqno{}  $$
where $C$ is the appropriate constant
establishing Weyl's law with a good error term for the noncompact space X.

The proof uses the Selberg trace formula in a form that is modified from the
work of Wallace and also draws on results of Stade and Wallace and
techniques of Huntley and Tepper. We also, in the course of the proof, give
an upper bound on the number of cusp forms that can violate the Ramanujan
conjecture.
\end{abstract}

\section{\bf{Introduction and statement of results}}

The study of the spectrum of the Laplacian on spaces has been a subject of
great interest in this century. This spectrum has applications to physics,
geometry and number theory. The first important result was obtained by Weyl
${\cite{We}}$, who showed that for a bounded region in $R^n\,\,\,$with
reasonable boundary conditions, the following holds 
$$
N(\lambda )\sim C\lambda ^{\frac {n}{2}}.
\eqno{}  $$
Here $N(\lambda )$ denotes the dimension of the space of eigenvalues of the
Laplacian with eigenvalue less than $\lambda $. Also,$\,C$ is a constant
that depends only on $n$, the dimension of the space. This result is known
as Weyl's law. It is, for many purposes, useful to have an error term for
the asymptotic result. For the case of bounded domains studied by Weyl,
Courant ${\cite{Co}}$ obtained the result 
$$
N(\lambda )=C\lambda ^{\frac {n}{2}}+O(\lambda ^{\frac{n-1}{2}}\log \lambda ).
\eqno{}  $$
The proofs of these results use the technique of Dirichlet-Neumann
bracketing, and the fact that the eigenvalues can be explicitly computed for
cubes.

For compact manifolds, Minakshisundaram and Pleijel ${\cite{M-P}}$ use
estimates on
the heat kernel to derive Weyl's result. Using the wave equation Hormander
${\cite{H}}$ obtains this result with the error term
$O\left( \lambda ^{\frac{n-1}{2}}\right) $. He does this by analyzing the
wave equation.

For noncompact manifolds Weyl's law is in general false. There is continuous
spectrum that does not allow the use of the above technique in their
original form. However, if the space is a finite volume symmetric space,
then other techniques are available. For symmetric spaces associated to 
$Sl(2,R)$ and a discrete group of isometries $\Gamma $ Selberg's trace
formula ${\cite{S}}$ can be used to obtain that Weyl's law is true if one
adds an
extra term that accounts for the continuous spectrum. Also, a reasonable
error term can be obtained. Moreover, if $\Gamma $ is a congruence group of 
$Sl(2,R)$ the additional term is smaller than the term accounting for the
square integrable eigenfunctions, so Weyl's law holds. Actually, slightly
more is true. We are now considering noncompact spaces, and so the spaces
have cusps. With a finite number of exceptions all of the square integrable
eigenforms are cusp forms, meaning that they vanish at the cusps. (An
equivalent definition involving the vanishing of certain integrals can be
given.)

The space that is of interest to us in this paper is 
$$
X=SL(3,Z)\backslash Sl(3,R)/SO(3,R).
\eqno{}  $$
One would expect for this space that similar results hold. Stade and Wallace
${\cite{S-W}}$, using the version of the trace formula for $X$ derived by
Wallace
${\cite{Wa}}$, show that for this space Weyl's law holds. The proof consists
of using
Gaussians as test functions to obtain the asymptotic behavior at small time
for the trace of the heat kernel. The result then follows from a standard
Tauberian Argument.

In this paper we show how this result can be combined with a different set
of test functions to obtain an error estimate for Weyl's law. We state this
explicitly.
\begin{description}
\item[Theorem 1.1:]  Let 
$$
X=Sl(3,Z)\backslash Sl(3,R)/SO(3,R)
\eqno{}  $$
Let $N(\lambda )$ denote the dimension of the space of cusp forms on $X$
with Laplace eigenvalue less than $\lambda$. Then 
$$
N(\lambda )=C\lambda ^{\frac {5}{2}}+O(\lambda ^2)
\eqno{}  $$
with 
$$
C=\frac{Vol(X)}{(4\pi )^{(\frac 52)}\Gamma ({\frac 72})}.
\eqno{}  $$
\end{description}

\smallskip

The constant given is the constant that one expects for Weyl's law. It
depends only on the dimension of the space, and the particular space. The
proof will be obtained by estimating the number of cusp forms with Laplace
eigenvalue between $\lambda $ and $\lambda +1$. We should note that cusp
forms are also eigenvalues of a third order differential operator as 
$Sl(3,R) $ has rank 2. We will not need this operator. Our test functions
will not be Gaussians, but rather the translations of functions that
approximate the characteristic function of a disc. In the course of the
proof we will give an estimate on the number of cusp forms that come from
representations that violate the Ramanujan conjecture, which in
representation theory language states that all cusp forms arise from
principal series representations. This set is conjectured to be empty, but
very little is known about it. In ${\cite{C-H-J-T}}$, it is shown that no
``small''
eigenvalues exist; however this does not prove the Ramanujan conjecture in
this case. For this reason it seems worthwhile to state this result as a
theorem.

\smallskip

\begin{description}
\item[Theorem 1.2:]  The dimension of the space of cusp forms with Laplace
eigenvalue less than $\lambda $ that violate the Ramanujan conjecture is 
$O(\lambda ^{\frac 32})$.
\end{description}

\smallskip

The Section 2 of this article will contain definitions and notations, and we
will write down the trace formula in a form that is useful for our purposes.
This is really just a changing of variables from the formula in
${\cite{S-W}}$. In Section 3 the theorems are proved.

\section{\protect\bigskip \bf{Notations and the Selberg Trace Formula}}

We will let $\lambda $ denote the eigenvalue corresponding to an 
$L^{2}$ eigenfunction of the Laplace-Beltrami operator on $X$. One may further
assume that the eigenfunction is actually a joint eigenfunction of the
entire ring of invariant differential operators, as that is a commutative
ring. We make no explicit use of the other independent operator (there are
two as $X$ has rank two), however when using the Selberg trace formula we
are implicitly using the fact we are working with automorphic forms, which
are joint eigenvalues of the entire ring. It is known that except for the
constant function, any such eigenfunction is a cusp form.

The fact that the eigenfunctions are actually eigenfunctions of two
operators makes it useful, and for the purpose of using the trace formula,
necessary, to write $\lambda $ in terms of two parameters. There are several
ways to do this, each with their own advantages. There is the representation
theory notation, in which the tempered representation, (those induced from
unitary characters, also those which would not violate the Ramanujan
conjecture, also those whose matrix coefficients are in $L^{2+\epsilon }$
for all $\epsilon >0$) are parametrized by two purely imaginary numbers. The
other representations are parametrized by pairs of complex conjugate numbers
with small, explicitly bounded, real part. As we will not need this we will
not go into more detail.

In ${\cite{S-W}}$ two parameters $s$ and $t$ are used to describe the
eigenvalues. In
this notation $\lambda =s(1-s)+{\frac {1}{3}}t(1-t).$ This notation works well
for
them as they write the trace formula in terms of Helgason transforms
${\cite{T}}$ and harmonic analysis on groups.

We, however, find it more convenient to work with a formula that relates a
function in two variables to its Euclidean Fourier transform. We then have
at our disposal all of the tools of classical harmonic analysis, and this
will make it easier to perform our analysis with a greater value of test
functions as we will not as in ${\cite{S-W}}$ have to explicitly evaluate
Fourier
transforms. We use parameters $\alpha $ and $\beta $ which are related to
the earlier parameters by $\alpha =-i(s-{\frac {1}{2}}),\beta
=-i(t-{\frac {1}{2}})$. We
also change the metric so that our $\lambda $ corresponds to $\frac {\lambda}
{3}$ in the notation of ${\cite{S-W}}$. This makes no real difference, as
this change
in the number of cusp forms will be accounted for by the change of volume
that it creates. In this notation we have$\,\lambda =3\alpha ^2+\beta ^2+1$.
Also, from representation theory it is known that either both $\alpha $ and 
$\beta $ are real or 
$-Im \sqrt{3}\alpha =Im \beta ,~~~\frac {1}{6}< Re \sqrt{3}\alpha
=Re \beta <\frac {1}{2}$.

It is also convenient to make the following change of notation from that in
${\cite{S-W}}$. On page 243, at the beginning of the second section, diagonal
matrices $a$ are introduced. They are diagonal matrices ,i.e. elements of
the group $A$ of diagonal matrices in $Sl(3,R)$. A given $a$ is described as 
$a=diag(a_{1,}a_2,(a_1a_2^{-1})$. Here we have $a_i>0$. Replacing $a_i$ by 
$e^{b_i}$ we obtain a pair of real numbers. We use this to replace the
Helgason transforms, that are essentially Mellin transforms in ${\cite{S-W}}$
by Fourier transforms.

We now state the trace formula in the form that we will use.

\smallskip

\begin{description}
\item[Proposition 2.1:]  Let $g$ be a smooth, compactly supported function
on $R^2$ such that its Fourier transform 
$\hat{g}$ 
is Weyl
invariant. Let ($s_n,t_n)$ denote the $(\alpha ,\beta )$ that correspond to
an eigenvalue $\lambda $ corresponding to a square integrable automorphic
form. Then 
$$
\sum\limits_{n=o}^\infty 
\hat{g}(s_n,t_n)=(ID)+(HYP)+(LOX)+(PAR)
\eqno{}  $$
\end{description}

where the 4 terms on the right are as follows. 
\begin{eqnarray*}
(ID) &=&\frac{volX}{(4\pi )^{\frac 52}\Gamma (\frac 72)}\int_{R^2}
\hat{g}(s,t)t(t-3s)(t+3s) \\
& & 
\tanh 
\pi t\tanh \frac{\pi (t+3s)}2\tanh \frac{\pi (t-3s)}2dsdt
\end{eqnarray*}

\begin{eqnarray*}
(HYP) &=&\sum_{\epsilon _1\epsilon _2\epsilon _3=1} Re[Z(\epsilon _1)]
Cl[Z(\epsilon _1)]\left| \epsilon _1^2\epsilon _2\right| \\
&&\lbrack (\epsilon _1-\epsilon _3)(\epsilon _1-\epsilon _2)(\epsilon
_2-\epsilon _3)]^{-1}g(\log \epsilon _1,\log \epsilon _2)
\end{eqnarray*}

$$
(LOX)=\sum_{(r,\theta )}\frac{\left| \ln r_0\right| Cl[Z(r)]}{1-2r^{-3}\cos
\theta +r^{-6}}\int_{R^2}\hat{g}(s,t)\frac{r^{1-s}e^{-2\theta 
Im~(t)}}{1+e^{-2\pi Im~(t)}}dsdt
\eqno{}  $$

$$
(Par)=A_1\hat{g}(0,0)+A_2\hat{g}(-\frac 16,\frac
12)+A_3\hat{g}(\frac 16,\frac 12)
\eqno{}  $$

We must explain the meaning of the terms. The spectral and the term (ID) are
self explanatory, and in (PAR) the $A_i$ are constants that need not be
computed for our purposes. For the term (HYP), the sum is over triples 
$\epsilon _1,\epsilon _2,\epsilon _3$ where the $\epsilon _i$ are real 
$\left| \epsilon _1\right| >\left| \epsilon _2\right|
>\left| \epsilon_3\right| $ 
and there is, by definition, a hyperbolic matrix in $Sl(3,Z)$
with $\epsilon _i$ as its eigenvalues. The $\epsilon _{i}$give rise
to a number field, and Reg denotes the regulator, and CL the narrow class
number. We also note that the logarithms occur in the function $g$ because
of our earlier change of variable $a=e^b$, that allowed us to work only with
Fourier transforms. Also the lack of a dependence on $\epsilon _{3}$
is illusory, as the product of the $\epsilon _i$ is 1. In the term (LOX),
the sum is over pairs $(r,\theta )$ with $r>0,r\neq 1,~~ 0<\theta \leq \pi $
such that some matrix in $Sl(3,Z)$ has eigenvalues $r^{-2},re^{i\theta
},re^{-i\theta }.$ If we denote the matrix by $g,$ then $r_0$ is such that
the centralizer of $g$ has generator conjugate to $diag(r_0^{-2},r_0e^{i
\theta },r_0e^{-i\theta })$. Finally, we noted that the test functions must
be Weyl invariant. This is a generalization of the fact that for the
classical trace formula, the test functions must be even. In our notation if
we let $\sqrt{3}s$ denote the $x$-axis and $t$ the $y$-axis, then two points
are related by the Weyl group if 
$$
(\sqrt{3}s_1,t_1)=(e^{\frac{n\pi i}3}\sqrt{3}s_2,e^{\frac{n\pi i}3}t_2)
\eqno{}  $$
This is simply a translation of representation theory facts into our
notation.

The proof of the proposition is simply a restating of Proposition 3.1
combined with proposition 4.1 in ${\cite{S-W}}$ and the change of variables
described above.

\section{\bf{Proof of Theorem}}

\smallskip Weyl's law has already been established; thus to prove theorem
1.1 it is sufficient to estimate the number of cusp forms with Laplace
eigenvalue between $\lambda -1$ and $\lambda $. We shall now proceed to do
this. Actually we will estimate the cusp forms with Laplace eigenvalue
between $\lambda -2\sqrt{\lambda }+1$ and $\lambda $. In the course of the
proof we will also prove Theorem 1.2.

We note that there are two ways that such an eigenvalue can exist. We can,
in our parameters, assume $S$ and $T$ are real and have 
$$
\lambda -2\sqrt{\lambda }+1\leq 3S^2+T^2\leq \lambda .
\eqno{}  $$
(Recall $\lambda =3\alpha ^2+\beta ^2+1$). We may also have $S$ and $T\,\,$
complex, and satisfying the restrictions given in Section 2, namely 
$$
-Im~~\sqrt{3}S=Im~~T
\eqno{}  $$
$$
\frac 16<Re~~\sqrt{3}S=Re~~T <~\frac 12
\eqno{}  $$
Actually, the complex parameters may be related to the above restriction by
a Weyl transformation as described in section 2.

We will handle the two situations separately, in both cases the trace
formula will be the main tool.

We first consider the case of real parameters. We will actually estimate the
number of eigenvalues with $\lambda -2\sqrt{\lambda }+1\leq 3S^2+T^2\leq
\lambda $. This is obviously sufficient. The point is that these
inequalities describe an annulus of inner radius $\sqrt{\lambda }-1$ and
outer radius $\sqrt{\lambda }$ if along the $x$-axis we put $\sqrt{3}S$ and
along the $y$-axis we put $T$.

We would like to put into the trace formula $\hat{g}$
equal to the characteristic function of the annulus. This is clearly not a
legitimate test function so we must approximate it. We start with a function 
$\hat{g}\,$such that $g_0$, its inverse Fourier transform, is
compactly supported, smooth, radially symmetric and nonnegative. This will
mean that $\hat{g}_0$ is Schwarz class and extends to an
analytic function of two complex variables with exponential growth. We may
also choose our functions so that 
$$
\hat{g}>0
\eqno{}  $$
$$
1<\hat{g}_0<2
\eqno{}  $$
if 
$$
3S^2+T^2<1
\eqno{}  $$
and 
$$
\hat{g}_0<\frac 2{(3S^2+T^2)^k}
\eqno{}  $$
for 
$$
3S^2+T^2>1
\eqno{}  $$
Here $k$ may be taken to be a large positive constant. We may also assume
that $\hat{g}_0>0$ for purely imaginary values of $S$ and $T$.
Standard arguments in Fourier analysis guarantee that such a function
exists.

For our test function, we choose translates of $\hat{g}_0$ such
that the point $(0,0)$ is translated to a point $(S_0,T_0)$ that is in the
annulus. We then sum over enough translates to cover the entire annulus.
This will take no more than the smallest integer greater than
$4\pi \sqrt{\lambda }$ 
translates. Here by cover,we mean that we want the test
function to be at least one in the annulus. By the positivity of $\hat{g}_0$ 
we see that if we estimate the contribution to the spectral
side of the equation for any given translate, we obtain an estimate for the
test function by summing the estimates of the translates. We now let 
($S_0,T_0)$ be a point in the annulus and we let 
$$
\hat{g}_{(S_0,T_0)}
\eqno{}  $$
be the translate of $\hat{g}_0$ that is centered at ($S_0,T_0)$, 
combined with its five Weyl transform related functions, so that it
approximates six characteristic functions.

\smallskip

\begin{description}
\item[Lemma 3.1:]  
$$
\sum\limits_{n=0}^\infty \hat{g}_{(S_0,T_0)}(s_n,t_n)=O(\lambda^{\frac 32})
\eqno{}  $$
when $s,t\in R$.
\end{description}

\smallskip

Remark: By summing over the $O(\sqrt{\lambda })$ such functions in the lemma
we will have that real $(S,T)$ contribute $O(\lambda ^2)$ to the spectral
side of the equation, proving part of the theorem.

Proof: We must obtain estimates for the terms on the right hand side of the
trace formula.

We first analyze the identity term. The contribution of this term with the
original function $\hat{g}_0$ is clearly $O(1)$. The translation
to the point $(\sqrt{3}S_0,T_0)$ will be $O(\lambda ^{\frac 32})$, as the
integrand is up to very small error, approximating by overestimating a cubic
polynomial multiplied by twice the characteristic function of a circle of
radius one. Of course, we are using the rapid decay of our test functions
outside of a circle of radius one.

The hyperbolic term involves the inverse Fourier transform of 
$$
\hat{g}_{({\sqrt{3}}S_0,T_0)}.
\eqno{}  $$
As 
$$
\hat{g}_{({\sqrt{3}}S_0,T_0)}
\eqno{}  $$
is just a translation of $\hat{g}_0$, this function will simply
be $g_0$ multiplied by a complex exponential. It is clear that from the
compact support of $g_0$ it gives an $O(1)$ contribution, as we only really
have a finite sum. The same will be true for such a function multiplied by a
complex exponential. We thus have an $O(1)$ contribution.

The loxodromic term is dominated in the annulus, up to some constant, by the
test function given in ${\cite{S-W}}$, when the Gaussian chosen is still
greater than 
$\frac 1{e}$ in the annulus. The decaying part of the test functions
can be easily estimated as we may choose $k$ in our original assumptions
arbitrarily large. We thus get a $O(\lambda )$ contribution for each 
$$
\hat{g}_{({\sqrt{3}}S_0,T_0)}
\eqno{}  $$

The parabolic terms give us $O(1)$ for each $\hat{g}_{({\sqrt{3}}S_0,T_0)}$ 
as we are evaluating at individual points, and as we translate
the function the result gets even smaller due to the rapid decay of the
original test function.

We thus have proved Lemma 3.1.

To complete the proof of Theorem 1.1 we will need to prove Theorem 1.2.

We now assume that $S$ and $T$ are not real, but satisfy the other possible
conditions, described above. We will once again use translates of $\hat{g}_0$. 
This time we will translate so that we cover the lines 
$T=0,~~T=3S,~~T=-3S$ up to the point $3\alpha ^2+\beta ^2+1<\lambda $. We will
need $O(\lambda ^{\frac 12})$ such translates. We once again use the trace
formula. For all terms but the identity, the arguments are identical to
those previously used. The identity term now gives a better estimate. We now
are only working with what is approximately a quadratic polynomial, as one
of $t,(t-3s),(t+3s)$ does not go to infinity as the parameters go to
infinity. We thus get an $O(\lambda )$ contribution from each 
$\hat{g}_{({\sqrt{3}}S_0,T_0)}$ yielding an over all contribution of 
$O(\lambda ^{\frac 32})$. This proves theorem 1.2 and thus we have proved
theorem 1.1. Once again, it should be emphasized that we actually expect
that the contribution from the nonreal $S$ and $T$ to only consist of the
constant function.

We conclude by remarking that a more refined analysis of the terms is likely
to be possible. This would allow us to use the techniques in ${\cite{H-T}}$
to prove
Weyl's law directly and to perform the refined local analysis found there,
leading to ''local'' Weyl's laws and more precise estimates. In particular,
our results show that ''on average'' every ball centered at
$({\sqrt{3}} S_0,T_0)$, 
with radius less than $\sqrt{\lambda }$ has approximately 
$C\lambda ^{\frac 32}$ where $C$ is a constant depending on the ball. Also
we expect to improve the estimates slightly.These issues will be discussed
elsewhere.

\begin{flushleft}
(1). Work supported in part by DOE contracts No. DE-AC-0276-ER 03074 and
03075; NSF Grant No. DMS-8917754; and PSC-CUNY
Research Award Nos. 6-6(6407,7418,8445).

(2),(3). Authors also acknowledge support for this research from the
Research Foundation of CUNY.  
\end{flushleft}


\newpage

\begin{thebibliography}{99}

\bibitem{We}  H. Weyl. Das Astmptotische Verteilungsgetz der Eigenwerte
linear partieller Differenmtialgleichungen. Math. Ann. 71(1911)441-469.

\bibitem{Co}  R. Courant. \"{U}ber die Eigenwerte bei dem
Differentialgleichungen der Mathematischen Physik. Math. Zeit. 7, 1920, 1-57.

\bibitem{M-P}  S. Minakshisundaram. A. Pleijel. Some properties of the
eigenvalue of the Laplacian. J. Diff. Geom. 1(1967), 43-69.

\bibitem{H}  L. Hormander. The spectral function of an elliptic operator.
Acta Math. 121(1968)193-218.

\bibitem{S}  A. Selberg. Harmonic analysis and discontinuous groups in
weakly symmetric Reimannian space with applications to Dirichlet series.  J.
Indian Math. Soc., 20(1956), 47-87.


\bibitem{S-W}  E. Stade, D. Wallace. Weyl's Law for $Sl(3,Z)\backslash
Sl(3,R)/So(3,R)$. Pacific J. of Math. Vol. 173, No. 1(1996) 241-261.

\bibitem{Wa}  D. Wallace. The Selberg Trace Formula for $Sl(3,Z)\backslash
Sl(3,R)/So(3,R)$, Trans. American Math. Society. Vol. 345, No. 1(1994) 1-36

\bibitem{C-H-J-T} S. Catto, J. Huntley, J. Jorgensen and D. Tepper.  On an
analogue of the Selberg's eigenvalue conjecture for $Sl(3,Z)$.
{\em Proc. Amer. Math. Soc. {\bf 126} (1998), pp.3455-3459.}

\bibitem{T}  A. Terras. Harmonic Analysis on Symmetric Spaces and
Applications. Springer-Verlag, New York, 1985.

\bibitem{H-T}  J. Huntley, D. Tepper. A local Weyl's law, the angular
distribution and multiplicity of cusp forms on product spaces. Trans. of
American Math. Society. 330 Number 1(1992)97-110.

\end{thebibliography}
\end{document}